\documentclass[lettersize,journal]{IEEEtran}
\usepackage{amsmath,amsfonts}
\usepackage{algorithmic}
\usepackage{algorithm}
\usepackage{array}
\usepackage[caption=false,font=normalsize,labelfont=sf,textfont=sf]{subfig}
\usepackage{textcomp}
\usepackage{stfloats}
\usepackage{url}
\usepackage{verbatim}
\usepackage{graphicx}
\usepackage{cite}
\usepackage{xcolor}

\begin{document}

\title{Numerical Differentiation-based Electrophysiology-Aware Adaptive ResNet for Inverse ECG Modeling 
}

\author{Lingzhen Zhu, Kenneth Bilchick, Jianxin Xie}



\maketitle

\begin{abstract}
Electrocardiographic imaging aims to noninvasively reconstruct the electrical dynamic patterns on the heart surface from body-surface ECG measurements, aiding the mechanistic study of cardiac function. At the core of ECGI lies the inverse ECG problem, a mathematically ill-conditioned challenge where small body measurement errors or noise can lead to significant inaccuracies in the reconstructed heart-surface potentials. 
To improve the accuracy of ECGI and ensure that cardiac predictions adhere to established physical principles, recent advances have incorporated well-established electrophysiology (EP) laws into their model formulations. However, traditional EP-informed models encounter significant challenges, including overfitting to EP constraints, limitations in network scalability, and suboptimal initialization. These issues compromise prediction accuracy and stability, hindering their effectiveness in practical applications. This highlights the need for an advanced data analytic and predictive tool to achieve reliable cardiac electrodynamic restoration. Here, we present a Numerical Differentiation-based Electrophysiology-Aware Adaptive Residual neural Network (EAND-ARN) for robust inverse ECG modeling. Our method employs numerical differentiation to compute the spatiotemporal derivative, enabling EP constraints to be applied across a local spatiotemporal region, thereby strengthening the overall EP enforcement. Additionally, we design an adaptive residual network to improve gradient flow, enhancing predictive accuracy and mitigating issues with poor initialization. Experimental results show that EAND-ARN significantly outperforms existing methods in current practice.
\end{abstract}


\section{Introduction}
The advancement of medical sensing and imaging technologies has created a data-rich environment, enabling smart health solutions that support more precise diagnostics and personalized treatment approaches. Among these advances, electrocardiography (ECG) stands out as a critical tool for non-invasive cardiac monitoring and diagnostics. It measures electrical signals on the body surface originating from the myocardium, capturing the spatiotemporal patterns that reflect the heart’s rhythmic activity. 
While the traditional twelve-lead ECG \cite{yang2013spatiotemporal} is effective for disease detection, it does not capture sufficient detail to provide comprehensive information about the heart's electrical distribution and spatial dynamics. 
Body Surface Potential Mapping (BSPM) extends ECG's utility by using a higher sensor density across the torso to refine cardiac electrical assessments and pinpoint dysfunction, providing the foundation for Electrocardiographic Imaging (ECGI) to noninvasively reconstruct heart surface potentials and enhance cardiac disease evaluation \cite{horacek2001optimal,bond2009xml,lacroix1991evaluation,rudy1999noninvasive,zhu2018optimal,chen2015sparse}.


While high-resolution BSPM provides comprehensive spatiotemporal information about cardiac electrical activity, translating this surface data into meaningful insights about the underlying cardiac electrodynamics requires solving the inverse ECG problem \cite{rudy1999noninvasive,oster1997noninvasive,gulrajani1998forward,yao2016mesh,yao2020spatiotemporal,macleod1998recent}. However, this problem is inherently ill-posed, as small errors or noise in the body measurements can lead to disproportionately large errors in the heart solution\cite{patel2022solution}. This challenge underscores the need for robust data analytics tools that can effectively handle the high-dimensional data structure and measurement noise management for inverse ECG modeling.

 Statistical regularization is a commonly used approach to stabilize the reconstruction of cardiac electrodynamics by suppressing unreliable variations in heart signals. For example, Tikhonov regularization\cite{ramanathan2004noninvasive,tikhonov1977solutions}, which applies a penalty with the L2 norm, promotes spatial smoothness in the reconstructed heart signal by reducing high-frequency noise. To further enhance coherence across time, spatiotemporal regularization (STRE) methods \cite{yao2016physics} have been developed, promoting not only spatial smoothness but also temporal continuity. However, these statistical regularization methods focus primarily on matching heart predictions with sensor observations, and their expressiveness is limited by model structures. Consequently, they often yield skewed and noisy cardiac solutions. 

In cardiac research, several phenomenological models have been developed to characterize the propagation of electrical waves across the heart, providing valuable physical insights into the mechanisms underlying cardiac function \cite{camara2011inter,sermesant2006electromechanical}. Solving these models merely provides theoretical representations of heart dynamics, without the ability to incorporate real-world sensing data. Consequently, these well-established Electrophysiology (EP) rules are often overlooked in the traditional regularization approaches. 
Recent research has advanced inverse ECG modeling by integrating EP laws into deep learning frameworks. 
This framework improves cardiac predictions by ensuring alignment with body ECG observations while adhering to established EP principles\cite{xie2022physics, sahli2020physics}.



Despite the aforementioned advances, the EP-constrained Deep Learning (EP-DL) model still faces significant challenges. 
On the one hand, while increasing network depth could theoretically enhance feature expressiveness and nonlinear feature extraction, prior work often resorts to smaller, shallower networks to avoid issues like network degradation and ineffective network scaling \cite{xie2022physics}. On the other hand, the conventional feedforward architecture in traditional EP-DL models may introduce initialization challenges \cite{wang2024piratenets},  increasing the risk of the network getting trapped in local minima during the early training stage. 
Furthermore, the EP-informed framework requires a large corpus of spatiotemporal collocation points to enforce the imposed EP rules during training, which introduces a substantial computational burden. Automatic differentiation (AD) calculates the derivatives in the EP model at individual spatiotemporal points, which does not consider neighboring relationships. This isolated differentiation approach may limit the model's ability to capture coherent spatiotemporal dynamics, leading to overfitting and suboptimal predictive accuracy. To address these limitations, we propose a Numerical Differentiation-based EP-aware Adaptive Residual Network (EAND-ARN) to improve the accuracy of inverse ECG solutions. Our major contributions are: 

1)We introduce an adaptive ResNet structure that mitigates initialization pathology, enabling stable and efficient scaling of the neural network architecture design for accurate cardiac predictions. 
Our proposed framework enhances traditional residual connections by introducing a trainable parameter constrained between 0 and 1. This setup facilitates the training of significantly deeper networks. 
Moreover, it effectively mitigates the poor initialization problem existing in the original EP-DL model, leading to more stable and accurate training outcomes.


2) We innovatively incorporate numerical differentiation to calculate the spatiotemporal derivatives when embedding the EP rules. To calculate the spatial Laplacian on a 3D heart geometry, we leverage a Laplacian operator derived from the spatial geometric relationships among neighboring nodes. Additionally, we utilize a fourth-order Taylor expansion to compute temporal derivatives based on the predicted signals at neighboring points. This approach enables the EP constraint to operate not only on individual instances but also to account for spatiotemporal locality, reducing the dependence on a large number of collocation points for enforcing EP rules.


Through experimental validation, our EAND-ARN framework demonstrates superior performance in inverse ECG reconstruction compared to state-of-the-art methods. Our results show significant improvements in reconstruction accuracy across different noise levels, marking an important advancement in precision cardiology and clinical applications.

The remainder of this paper is organized as follows: Section II provides a literature review of relative techniques to solve inverse ECG problems. Section III presents our EAND-ARN methodology, detailing the mathematical formulation of EAND-ARN architecture. Section IV provides experimental results and comparative performance analysis. Finally, Section V concludes with a summary of our findings.

\section{Literature Review}
\subsection{Traditional Methods for Solving Inverse ECG Problem}

The inverse ECG problem aims to non-invasively reconstruct the electrical activity on the heart surface using body surface sensor data. 
$\boldsymbol{R}$. 
Mathematically, this relationship can be expressed as:
$$y(s, t) = \boldsymbol{R} u(s, t) +\epsilon$$
where $u$ and $y$ denote the heart surface potential (HSP) and the body surface potential (BSP), respectively. $s$ and $t$ represent the spatial location and time instance. $\boldsymbol{R}$ is the transfer matrix that can be derived by the Divergence Theorem and Green’s Second Identity \cite{yao2016physics,barr1977relating}. $\epsilon$ stands for the noise term that captures errors from both measurement and modeling. The inverse ECG problem is inherently ill-posed due to the dimensional mismatch between the high-dimensional cardiac source distribution, representing the HSP to be estimated, and the limited number of body surface sensor measurements. This mismatch randers the transfer matrix $\boldsymbol{R}$ rank deficient and characterized by a high condition number. As a result, even minor uncertainties in body surface measurements are dramatically amplified when solving for the HSP.\cite{patel2022solution}. Accurate inverse ECG solutions are vital for ECGI systems, as they are essential for clinical applications such as localizing ectopic points and evaluating post-surgical outcomes. \cite{yadan2023expert}.

Many studies in the literature employ regularization methods to stabilize the inverse solution. 
A commonly used method is Tikhonov regularization, which stabilizes the inverse ECG problem by penalizing the unreliable magnitude of the solution using L2-norm constraints. The objective function for Tikhonov regularization can be expressed as:

\begin{align*}
    \min_{u(s,t)} \{\|y(s,t) - Ru(s,t)\|_2^2 + \lambda_{T}^2\|\Gamma u(s,t)\|_2^2\}
\end{align*}
where $\lambda_{T}$ the regularization parameter that controls the trade-off between data fidelity and smoothness and $\Gamma$ is the smoothing operator to promote the regularity in the HSP solution. Popular choices for $\Gamma$ in Tikhonov regularization include the identity matrix ($\Gamma = I$), which penalizes the magnitude of 
$u(s,t)$, and differential operators such as the gradient or Laplacian, which penalize the roughness or non-smoothness of the solution \cite{tikhonov1977solutions,wang2009resolution,kara2019ecg,bear2020impact,schuler2021reducing,rababah2021effect}. 
However, Tikhonov regularization often oversmooths the solution, potentially obscuring important features. Additionally, it processes each time instant independently, failing to account for temporal dependencies in cardiac electrical activity. 

To better incorporate temporal information, spatiotemporal regularization techniques were developed. Messnarz et al. \cite{messnarz2004new} introduced a spatiotemporal approach that combines first-order Tikhonov regularization in the spatial domain with a temporal constraint that assumes non-decreasing potential values during the depolarization phase. This method improves reconstruction stability but has limitations in handling abnormal cardiac conditions where the non-decreasing assumption may not hold. Yao et al.  \cite{yao2016physics} proposed the STRE model, which introduces a spatial Laplacian operator to handle approximation errors through spatial correlations on complex geometries, while simultaneously applying temporal constraints over specified time windows to increase model robustness to measurement noise. 
However, STRE is computationally intensive and does not incorporate the well-established EP rules. 

Recent advances have integrated EP rules to enhance the reconstruction of HSP. For example, Wang et al. \cite{wang2009physiological} developed a physiological-model-constrained Kalman filter, which simulates the cardiac system using a high-dimensional stochastic state-space model, reconstructed heart potential distributions using an adapted unscented Kalman filter. However, the accuracy of their HSP prediction heavily relies on initial electrical potential map on the heart surface. Jiang et al. \cite{jiang2024hybrid} developed a hybrid neural state-space model to address the inverse ECG problem by leveraging deep neural networks to learn the Bayesian filtering and transition functions. However, this model does not explicitly encode fundamental EP laws; instead, it entirely relies on DNN to approximate the underlying dynamic patterns. 
Our prior work \cite{xie2022physics} utilized deep neural networks (DNNs) as approximators, incorporating EP rules into the loss function to impose EP constraints. While the model demonstrated promising results, it was limited by a shallow network architecture and exhibited potential overfitting to the imposed EP constraints. As such, there remain challenges in effectively enforcing the EP rules and optimizing the DNN architectures. 

\subsection{Physics-informed Neural Networks (PINNs)}

Physics-informed neural networks (PINNs) have emerged as a powerful framework for studying complex physical systems by embedding the physical laws into the model design. 
 Since their formalization and popularization by Raissi et al. \cite{raissi2019physics}, PINNs have been successfully applied across various domains, including fluid dynamics \cite{cai2021physics}, solid mechanics \cite{rao2021physics}, and biomedical engineering \cite{xie2022physics, sahli2020physics, herrero2022ep, sarabian2022physics, zhang2023physics}. By leveraging governing equations, such as partial differential equations (PDEs), as soft constraints within the loss function, PINNs have demonstrated the ability to solve forward and inverse problems with remarkable efficiency and accuracy.

Building on their success across various fields, PINNs have been increasingly recognized in the field of cardiac electrophysiology. Cardiac electrophysiology studies the electrical processes that govern the heart's rhythm and coordination, driving the contraction of cardiac muscle to pump blood throughout the body. The knowledge of cardiac electrodynamics is crucial for diagnosing, predicting, and treating cardiac arrhythmias and other electrophysiological disorders \cite{thangaraj2024cardiovascular}. As such, scientists have been studying the cardiac dynamic mappings by incorporating the well-developed EP laws into PINNs. For example,
Martin et al. \cite{herrero2022ep} demonstrated the capability of PINNs to reconstruct action potential propagation in 1D and 2D grid structure using the Aliev-Panfilov (AP) model. Sahli et al. \cite{sahli2020physics} reconstruct cardiac activation mappings in 2D grid by accounting for both underlying wave propagation dynamics characterized by Eikonal equation and the observation data from the cardiac surface. However, these methods did not account for the complex 3D geometry of the heart. Xie et al. \cite{xie2022physics} proposed an EP-DL framework that integrated the AP model into deep neural networks to study the surface electrical dynamics from body surface potential mapping (BSPM), showing good reconstruction accuracy and robustness against noise. Ye et al. \cite{ye2023spatial} developed an adaptive PINN framework for bi-ventricular electrophysiological processes using a volumetric heart model that accounts for myocardial thickness. Their approach incorporates the heart’s initial states to improve accuracy, whereas in real-world scenarios, such information is typically unavailable. 

Moreover, the above-mentioned PINN implementations rely on automatic differentiation (AD), which computes exact derivatives through chain rule during backpropagation \cite{baydin2018automatic}. Although AD provides exact gradients, this approach can lead to nonphysical solutions by perfectly satisfying equations at isolated collocation points while missing broader spatial patterns \cite{chiu2022can}. When neural networks are heavily over-parameterized, AD-formulated loss functions often become under-constrained optimization problems \cite{sirignano2018dgm}. This limitation is particularly problematic in systems requiring high-dimensional spatial representations. 
In contrast, numerical differentiation (ND) approximates derivatives using neighboring points \cite{anderson2020computational}, ND can address the limit by naturally incorporating neighboring points through local support regions, allowing PINNs to learn coherent patterns even with sparse sampling of collocation points \cite{chiu2022can}. 

In addition, traditional PINNs for simulating cardiac electrodynamics encounter significant training challenges, including poor initialization, where the magnitude of the EP loss term dominates the data-driven loss at the beginning epochs. This imbalance increases instability during training, leading to slow convergence and suboptimal results \cite{cuomo2022scientific, wang2024piratenets}. Additionally, feed-forward neural networks, commonly used in PINNs, face gradient vanishing issues and performance degradation as network depth increases, limiting their capacity to model complex spatiotemporal dynamics in deep architectures \cite{berg2018unified}. These challenges highlight the need for innovative neural network architectures that maintain stability and enhance performance as the network depth increases.

\section{Methodology}
\subsection{Cardiac Electrophysiology Model}
Cardiac electrophysiology plays a pivotal role in understanding the dynamics of electrical signal propagation in the heart.
To achieve a balance between computational efficiency and the ability to study critical cardiac dynamics, we adopted the Aliev-Panfilov (AP) model \cite{aliev1996simple}, a commonly used framework for simulating cardiac excitation and signal propagation in cardiac tissue. The AP model is defined by the following set of partial differential equations: 

\begin{align}
    \frac{\partial u}{\partial t} &= \nabla \cdot (D\nabla u) u + k u (u - a)(1 - u) - u v  \label{Eq: AP1}\\
    \frac{\partial v}{\partial t} &= \xi(u, v)(-v - k u (u - a - 1)) \label{Eq: AP2}\\
    \mathbf{n} \cdot \nabla u \big|_{\Gamma_H} &= 0 \label{Eq: bd}
\end{align}
where $u$ represents the normalized HSP, and $v$ denotes the recovery current that governs the local depolarization of membrane potential. The coupling interaction between $u$ and $v$ is characterized by $\xi(u, v) =  e_0 + \mu_1 v / (u + \mu_2)$. $D$ is the diffusion coefficient that controls the conduction velocity, $k$ is the repolarization constant that controls the shape of the action potential, and $a$ controls the tissue excitability. In this study, we assume the electrical conductive homogeneity of the heart tissue, which allows us to simplify the diffusion term as $\nabla \cdot (D\nabla u)=D\Delta u $, where $\Delta u$ represents the Laplacian of the HSP $u$. The values of these parameters are set based on existing literature \cite{aliev1996simple}: $a = 0.1$, $D = 10$, $k = 8$, $e_0 = 0.002$, and $\mu_1 = \mu_2 = 0.3$. Eq. \ref{Eq: bd} is a Neumann boundary condition that ensures there is no flux of the electrical potential $u$ across the heart boundary $\Gamma_H$, where $\mathbf{n}$ is the unit normal vector to the boundary $\Gamma_H$. The AP model defined in Eqs. \ref{Eq: AP1}-\ref{Eq: bd}, will be integrated into a deep learning framework to further enhance the HSP prediction, ensuring that the results are grounded in physical significance.

\subsection{EP-Aware Deep Learning Approach}


\begin{figure*}[htp]
\centering
\includegraphics[width=6.5in]{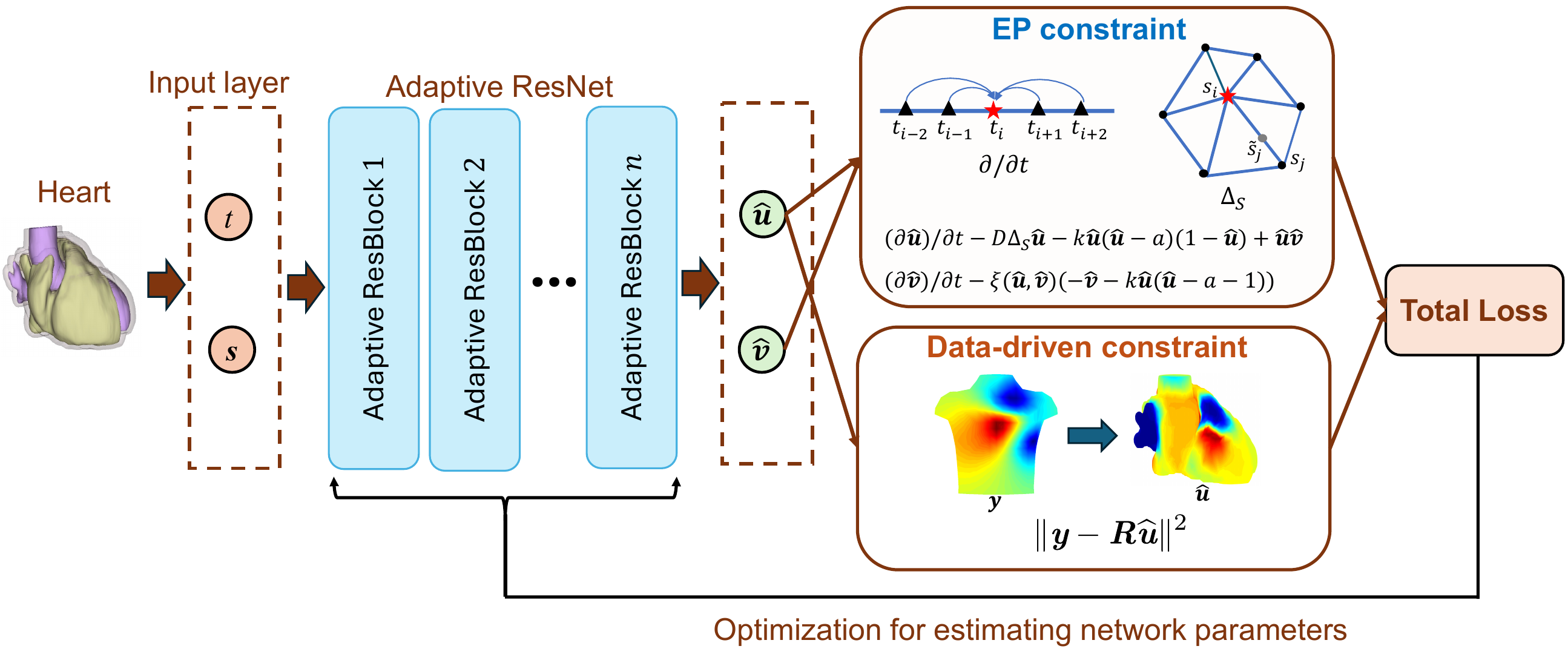}
\caption{Illustration of the proposed EAND-ARN framework for cardiac electrodynamic prediction. }
\label{Fig: Framework}
\end{figure*}

The EAND-ARN framework of our proposed model is shown in Fig. \ref{Fig: Framework}. The inverse ECG solution is parameterized by a deep learning network (DNN). Specifically, the input of the DNN consists of the spatiotemporal coordinates $[s, t]$, where $s$ denotes the discretized 3D node on the heart geometry and $t$ is the temporal instance. The DNN serves as a nonlinear functional approximator to project the spatiotemporal instances $[s,t]$ into the prediction of the heart signals--HSP $u(s, t)$ and recovery current $v(s, t)$. In our proposed framework, the AP model is integrated into the DNN through a customized loss function. This loss function takes both body sensor observation and the EP rules into consideration, ensuring spatiotemporal predictions of HSP that align with BSPM measurements while adhering to the physiological principles of cardiac electrophysiology: 
\begin{equation}
    \mathcal{L}_{\text{total}} = \mathcal{L}_{\text{D}} + \lambda \cdot \mathcal{L}_{\text{EP}}
    \label{Eq: loss_total}
\end{equation}
where $\mathcal{L}_{\text{total}}$ represents the total loss, $\mathcal{L}_{\text{D}}$ and $\mathcal{L}_{\text{EP}}$ represent the data-driven loss and EP-aware loss respectfully, $\lambda$ is a hyperparameter to regulate the intensity of EP-aware loss. The selection of $\lambda$ is shown in Section \ref{section: lambda}. The two major losses $\mathcal{L}_{\text{D}}$ and $\mathcal{L}_{\text{EP}}$ are introduced below:


\textit{1) Data-driven Loss $\mathcal{L}_{\text{D}}$:} The data-driven loss $\mathcal{L}_{\text{D}}$ is designed to match the predicted HSP $\hat{u}$ with the body sensor observation $y$. The relationship between the HSP and BSPM is described by a forward model, where the BSPM values $y\in \mathbb{R}^m$ are linearly related to the HSP values $u\in \mathbb{R}^n$ through a transfer matrix $\boldsymbol{R}\in \mathbb{R}^{m\times n}$,  such that $y = \boldsymbol{R}u$, where the derivation of transfer matrix $\boldsymbol{R}$ can be referred to \cite{yao2020spatiotemporal,barr1977relating}. For a given HSP prediction $\hat{u}$, the corresponding predicted BSPM $\hat{y}$ is obtained as $\hat{y} = \boldsymbol{R} \hat{u}$.  The data-driven loss function $\mathcal{L}_\text{D}$ is then computed to minimize the difference between body signal estimation 
$\hat{y}$ and the actual BSP observation $y$, ensuring the model predictions respect the observed data:


\begin{equation}
    \mathcal{L}_{\text{D}} = \frac{1}{N} \sum_{t} \sum_{s} \| y(s,t) - \hat{y}(s,t) \|^2
\end{equation}

where $N$ stands for the total number of the spatiotemporal instances. Due to the rank deficiency of the transfer matrix $\boldsymbol{R}$, the inverse solution $\hat{u}$ is usually ill-posed. As such, an effective EP rule grounded regularization should be posed to increase the prediction accuracy. 

\textit{2) EP-Aware Loss:} To further improve the accuracy and robustness of the heart prediction, we impose the AP model-based constraints on a set of spatiotemporal collocation points $\{[s_k,t_k]\}|_{k=1}^{N_c}$'s, which are randomly selected among the cardiac spatiotemporal domain. $N_c$ is the total number of the selected collocation points. The EP rule is embedded by penalizing the HSP predictions $\hat{u}$ that deviates from the AP model. As such, we can be mathematically  expressed in the form of residuals derived from Eq. \ref{Eq: AP1}-\ref{Eq: bd}. Specifically, the residuals can be written as:



\begin{align}
    r_u(s, t) &= \frac{\partial \hat{u}}{\partial t} - D \cdot\Delta \hat{u} - k_r \hat{u}(1 - \hat{u})(\hat{u} - a) + \hat{u} \hat{v}\\
    r_v(s, t) &= \frac{\partial \hat{v}}{\partial t} - \varepsilon(\hat{u}, \hat{v})(-\hat{v} - k_r \hat{u} (\hat{u} - a - 1))\\
    r_b(s,t) &= \mathbf{n}\cdot \nabla \hat{u}(s,t), \qquad s\in \Gamma_H \nonumber \nonumber\\
\end{align}

The EP injection will be accomplished by minimizing the magnitudes of the resdiuals $r_b(s,t)$, $r_u(s,t)$, and $r_v(s,t)$ at the selected collocation points. Hence, the EP-aware loss is defined as:


\begin{equation}
    \mathcal{L}_{\text{EP}} = \frac{1}{N_c} \sum_{k=1}^{N_c} \left[ r_u(s_k, t_k)^2 + r_v(s_k, t_k)^2 + r_b(s_k, t_k)^2\right]
\end{equation}

By training the DNN integrating the data-driven loss $\mathcal{L}_\text{D}$ with the EP aware loss $\mathcal{L}_{\text{EP}}$, we expect that the HSP prediction $\hat{u}$ will not only comply to the body sensor observation $y$ but also stick to the EP rule. 

It is worth noting that the estimation of the derivatives $(\frac{\partial \hat{u}}{\partial t}, \frac{\partial \hat{v}}{\partial t}, \Delta \hat{u})$ is critical to accurately enforcing the EP rule and ensuring that the model predictions adhere to the underlying EP. Hence, employing an effective strategy to compute these derivatives is essential to mitigate overfitting to the sampled collocation points. Moreover, the DNN with deepened depth and stable training process also marks an important component in achieving higher accuracy of HSP prediction. As such, we engage a numerical differentiation to impose the EP rule, not merely on single points, but at local spatiotemporal region, thus tightening the enforcement of EP. Meanwhile, we propose an adaptive residual neural network to broaden the network structure selections, aiming to further enhance the HSP predictive accuracy.

\subsection{Adaptive Residual Neural Network}
Multilayer perceptrons (MLPs) are the most commonly used neural network architecture in PINNs. \cite{saravanan2014review,abiodun2018state,wang2018deep}. A MLP is a type of artificial neural network composed of an input layer, one or more hidden layers, and an output layer, where each layer is fully connected to the next. Each neuron in an MLP processes a weighted sum of its inputs and passes the result through an activation function, typically non-linear, enabling the network to learn complex patterns and approximate highly non-linear functions. The output of the $l$-th layer can be mathematically expressed in its general form as:


\begin{equation}
    \mathbf{x}^{(l)} = \sigma(\mathbf{W}^{(l)} \mathbf{x}^{(l-1)} + \mathbf{b}^{(l)})
\end{equation}

where $\mathbf{x}^{(l-1)}$ represents the output from the $(l-1)$-th layer, $\mathbf{W}^{(l)}$ is the weight matrix, $\mathbf{b}^{(l)}$ is the bias vector for $l$-th layer, and $\sigma(\cdot)$ is the activation function. Their ability to effectively model non-linear relationships makes MLPs well-suited for studying non-linear physics systems. 

Previous research has demonstrated the effectiveness of MLPs in solving inverse ECG problems \cite{xie2022physics, sahli2020physics, herrero2022ep}. However, it is found that the trainability of the EP-DL introduces ineffective scaling or even degradation as the MLP network increases. As such, EP-DL frameworks \cite{xie2022physics} for studying the cardiac dynamic signals typically employ small and shallow neural networks with only 5 layers, limiting their ability to fully leverage the advantages of deeper architectures. 
Moreover, conventional EP-DL approaches often face challenges with pathological initialization, where the initial training epochs produce an unbalanced loss with large magnitudes, resulting in unstable optimization and slower convergence. 

\begin{figure}[!t]
\centering
\includegraphics[width=3.0in]{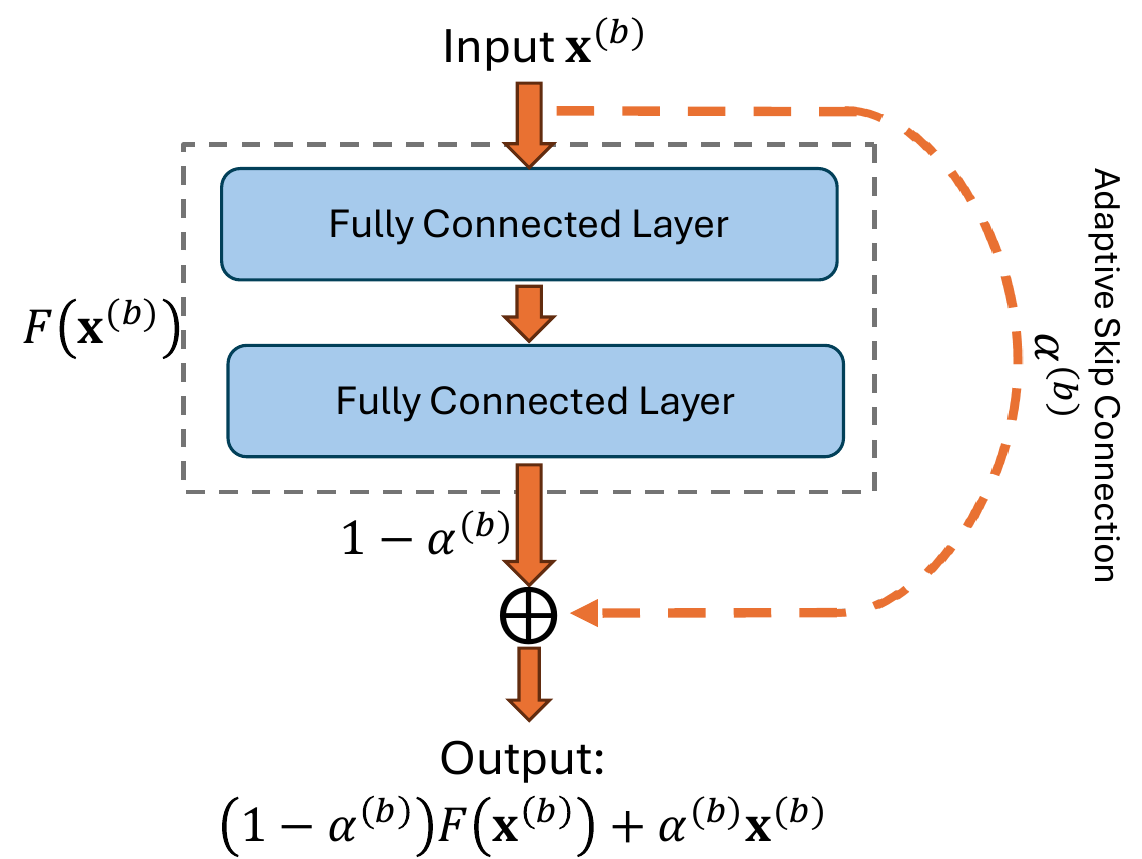}
\caption{Illustration of Adap-ResBlock.}
\label{resblock}
\end{figure}

To overcome the above limitations, our study innovatively introduces a novel adaptive residual block, specifically designed to address the problem of ineffective scaling and ill-posed initialization problem. 
The architecture of the proposed adaptive residual block is shown in Fig. \ref{resblock}. Specifically, we defined a shortcut connection across stacked layers, allowing the input $x$ to flow directly to the output. The mathematical representation of adaptive residual block is defined as:


\begin{align}
\mathbf{h_1}^{(b)} &= \sigma(\mathbf{W_1}^{(b)} \mathbf{x}^{(b)} + \mathbf{b_1}^{(b)}) \nonumber\\
\mathbf{h_2}^{(b)} &= \sigma(\mathbf{W_2}^{(b)} \mathbf{h_1}^{(b)} + \mathbf{b_2}^{(b)}) \nonumber\\
\mathbf{x}^{(b+1)} &= (1-\alpha^{(b)}) \mathbf{h_2}^{(b)} + \alpha^{(b)} \mathbf{x}^{(b)}
\end{align}

where $\mathbf{x}^{(b)}$ is defined as the input of the $b$-th adaptive residual block (Adap-ResBlock), $\mathbf{h_1}^{(b)}$ and $\mathbf{h_2}^{(b)}$ are the output of the hidden layers within the $b$-th Adap-ResBlock. $\mathbf{W}$'s and $\mathbf{b}$'s are the trainable neural net parameters. $\alpha^{(b)}$ is a hyperparameter that controls the proportion of the knowledge that is directly inherited from the input $\mathbf{x}^{(b)}$ at $b$-th Adap-ResBlock. $\sigma(\cdot)$ stands for the activation function. We choose the hyperbolic tangent function due to its general effectiveness in physics-informed system learning \cite{jagtap2020conservative}.


To improve the explainability of the results, it is beneficial to determine the percentage of the input that flows directly to the output versus the percentage of the output contributed by the functional block. $\alpha^{(b)}$ is constrained between 0 and 1 to ensure that it represents a meaningful proportion. 
This is achieved by introducing a trainable parameter 
$\alpha_T^{(b)}$ and applying a sigmoid function, which ensures $\alpha^{(b)}$ stays within the desired range:  

\begin{equation*}
    \alpha^{(b)}= \frac{1}{1 + e^{-\alpha_T^{(b)}}}
\end{equation*}

By progressively deepening the network during training, these adaptive connections allow for an improved and more stable training process, making the network less susceptible to poor initialization and effectively mitigating the convergence issue. With the inclusion of adaptive residual blocks, we were able to add additional layers and neurons, further enhancing the model's ability to capture the complicated spatiotemporal dependencies in the HSP data. 



\subsection{Numerical Differentiation on Spatiotemporal Domain}

In the EP-DL framework, the accuracy of predicted cardiac electrodynamics is determined by the neural network parameters 
$\boldsymbol{W}_\theta$, optimized by minimizing the loss function $\mathcal{L}_\text{total}$ defined in Eq. 21. Among the components of the loss function, 
$\mathcal{L}_\text{EP}$ serves as a physics-based constraint, ensuring the predictions adhere to the AP model. Consequently, temporal derivatives ($\frac{\partial \hat{u}}{\partial t}$, $\frac{\partial \hat{v}}{\partial t}$) and the spatial Laplacian ($\Delta \hat{u}$)—play a critical role in constructing the EP-based loss, which is essential for correctly updating $\boldsymbol{W}_\theta$.

Traditionally, AD is employed to compute these differential operators precisely at specified collocation points, automatically generating accurate derivatives during backpropagation \cite{baydin2018automatic}. However, AD can achieve near-zero training loss by overfitting to collocation points, but this does not guarantee accuracy in sparse sampling regimes, making the loss metric unreliable without ground truth verification. As such, the effectiveness of the EP-aware approach relies on enforcing the constraints over a large number of residual points, making the training process computationally intensive \cite{fang2021high, lim2022physics}. 

Moreover, since all collocation points are individually constrained by the EP laws, AD inherently lacks the ability to incorporate information from the neighborhood within the domain. This limitation undermines the spatial coherence that is critical for accurately capturing the dynamics of electrical wave propagation. Additionally, because the electrical wave propagates across a complex 3D heart geometry, traditional finite difference methods (FDM), which rely on structured, grid-like discretization, are impractical for directly addressing computations on an irregular 3D surface.


To address the aforementioned challenges, we propose an ND-based approach for calculating the spatial Laplacian, which considers not only the HSP at the central spatial point but also its neighboring locations. Additionally, we employ a fourth-order Taylor expansion to compute the temporal derivatives. These two measures collectively enhance spatiotemporal coherence and improve the accuracy of the EP constraint, ensuring more reliable predictions of cardiac electrodynamics. The methods we propose are elaborated as follows:

\begin{itemize}
\item[1).] \textit{Spatial Laplacian}: 
  We propose an innovative approach to tighten the EP constraint given a pre-defined number of collocation points. We calculate the spatial Laplacian ($\Delta u$) using a neighbor-aware strategy that considers not only the values at the central point but also the contributions from its neighboring locations. 

In a regular 2D square grid shown in Fig. \ref{fig:laplacian}(a), the Laplacian at the center node $s_0$ can be derived given the signal value at $s_0$ and its four connected neighbors ($s_1, s_2, s_3, s_4$) using the FDM with a second-order Taylor expansion, yielding the Laplacian at node $s_0$ as 
\begin{equation}
\Delta u_0 = \frac{4}{d^2}(\bar{u}-u_0) 
\label{Eq: regular_grid}
\end{equation}
where $\bar{u}$ is the average of the values at the neighboring points, and $d$ is the distance between $s_0$ with the neighbors. While originally developed for regular grids, this FDM-based approach can be adapted for triangular meshes by redefining $\bar{u}$ to account for irregular geometries, enabling its application to 3D heart surfaces \cite{yao2016physics}.


\begin{figure}[hb]
\centering
\includegraphics[width=3.5in]{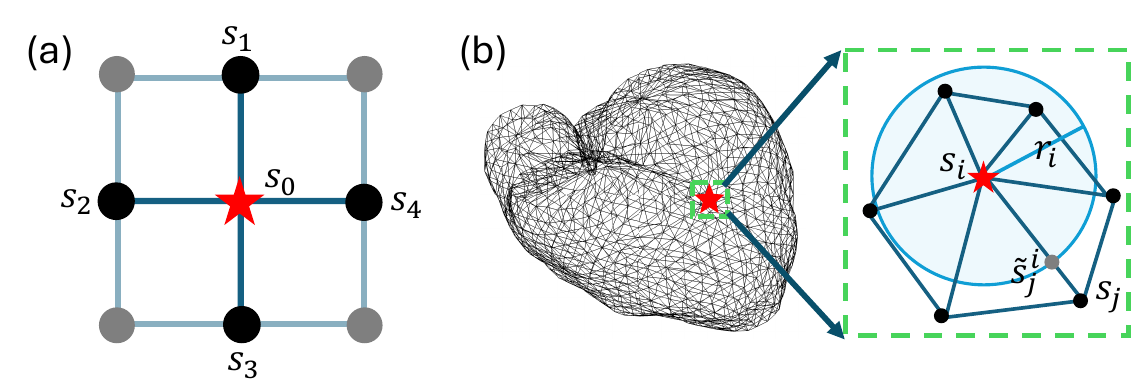}
\caption{(a) 2D square grid; (b) 3D triangle mesh on human heart surface}
\label{fig:laplacian}
\end{figure}


Specifically, the cardiac surface is discretized into triangle meshes, as shown in Fig. \ref{fig:laplacian}(b). By zooming in on a single node $s_i$ on the surface, its local neighborhood can be defined by the triangular connections formed with surrounding points. Let $u_i$ be the value of the HSP at $s_i$, and $d_{ij}$ be the distance from node $s_i$ to one of the neighbor nodes $s_j$. And $r_i$ is the average distance between all the neighbor points with $s_i$. $\tilde{s}_j^i$ represents a point located between the central node $s_i$ and its neighboring node $s_j$, positioned at a distance $r_i$ away from $s_i$. The HSP magnitude at $\tilde{s}_j^i$, denoted as $\tilde{u}_j^i$, can be computed using linear interpolation:

\begin{equation}
    \tilde{u}_j^i = u_i + \frac{r_i}{d_{ij}} (u_{j}-u_{i})
    \label{Eq: linear_interpolation}
\end{equation}

Each edge connection between $s_i$ with its neighboring point will have an intersection node. As such, when calculating the spatial Laplacian $\Delta_s(u_i)$ for irregular triangular mesh geometry, the $\bar{u}$ in Eq. \ref{Eq: regular_grid} can be replaced with the mean magnitude of the intersection nodes :
\begin{equation}
    \Delta_s(u_i) = \frac{4}{r_i}(\frac{1}{n_i} \sum_{j=1}^{n_i}\tilde{u}_j^i-u_i )
\end{equation}
where $n_i$ is the total number of the neighboring points. By substituting $\tilde{u}_j^i$ with the definition in Eq.\ref{Eq: linear_interpolation}, the final FDM-based spatial Laplacian at $s_i$ can be formulated as \cite{yao2016physics}:

\begin{equation}
\Delta_{s}\left(u_i\right)=\frac{4}{r_i n_i}\left(\sum_{j=1}^{n_i} \frac{u_j}{d_{i j}}-\sum_{j=1}^{n_i} \frac{u_i}{d_{i j}}\right)
\end{equation}




\item[2).] \textit{Temporal Derivative}: 
In the AP model, the temporal derivatives ($\frac{\partial \hat{u}}{\partial t}$, $\frac{\partial \hat{v}}{\partial t}$) are critical components for the EP-based regularization. However, AD computes these derivatives at individual temporal instances without accounting for the neighboring temporal context, which increases the risk of overfitting to the sparsely chosen temporal collocation points. Here, we tackle this challenge by employing the concept of FDM. 
We propose to leverage  higher-order approximation 
leverage a broader temporal context, achieving improved accuracy in estimating temporal derivatives.


\begin{figure}[!t]
\centering
\includegraphics[width=2.8in]{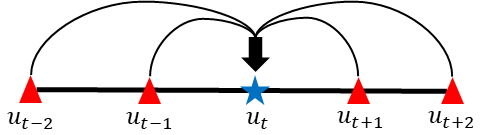}
\caption{The temporal derivative at $u_t$ derived from four temporal neighbors.}
\label{Fig: temporal_derivative}
\end{figure}

Specifically, we incorporate a fourth-order Taylor expansion-based FDM to approximate the first-order temporal derivative $\frac{\partial u}{\partial t}$. As shown in Fig. \ref{Fig: temporal_derivative}, this approach incorporates the HSP magnitudes from temporal neighboring points at $t\pm \tau$ and $t\pm 2\tau$, where $\tau$ denotes the time interval. The Taylor expansion at these points are given as:


\begin{align}
    u_{t-2\tau} &= u_{t} - 2\tau \frac{\partial u_{t}}{\partial t} 
    + \frac{(2\tau)^2}{2!} \frac{\partial^2 u_{t}}{\partial t^2} 
    - \frac{(2\tau)^3}{3!} \frac{\partial^3 u_{t}}{\partial t^3} \notag \nonumber \\
    &\quad + \frac{(2\tau)^4}{4!} \frac{\partial^4 u_{t}}{\partial t^4} 
    + O(\tau^5) \nonumber\\
    u_{t-\tau} &= u_{t} - \tau \frac{\partial u_{t}}{\partial t} 
    + \frac{\tau^2}{2!} \frac{\partial^2 u_{t}}{\partial t^2} 
    - \frac{\tau^3}{3!} \frac{\partial^3 u_{t}}{\partial t^3} + \frac{\tau^4}{4!} \frac{\partial^4 u_{t}}{\partial t^4} 
    + O(\tau^5) \nonumber\\
    u_{t+\tau} &= u_{t} + \tau \frac{\partial u_{t}}{\partial t} 
    + \frac{\tau^2}{2!} \frac{\partial^2 u_{t}}{\partial t^2} 
    + \frac{\tau^3}{3!} \frac{\partial^3 u_{t}}{\partial t^3} + \frac{\tau^4}{4!} \frac{\partial^4 u_{t}}{\partial t^4} 
    + O(\tau^5)\nonumber\\
    u_{t+2\tau} &= u_{t} + 2\tau \frac{\partial u_{t}}{\partial t} 
    + \frac{(2\tau)^2}{2!} \frac{\partial^2 u_{t}}{\partial t^2} 
    + \frac{(2\tau)^3}{3!} \frac{\partial^3 u_{t}}{\partial t^3} \notag \nonumber\\
    &\quad + \frac{(2\tau)^4}{4!} \frac{\partial^4 u_{t}}{\partial t^4} 
    + O(\tau^5)
\end{align}

where the value of $u_{t-2\tau}$, $u_{t-\tau}$, $u_{t+\tau}$ and $u_{t+2\tau}$ is known from the forward prediction $\hat{u}$ in each epoch. By combining these equations and canceling higher-order terms, we derive the fourth-order accurate formula for the temporal derivative:

\begin{align}
    \frac{\partial \hat{u}_c}{\partial t}\approx \frac{\hat{u}_{t-2\tau} - 8\hat{u}_{t-\tau} + 8\hat{u}_{t+\tau} - \hat{u}_{t+2\tau}}{12\tau}
    \label{Eq:temporal_central}
\end{align}


where $u_c$ denotes the HSP value at central points. Note that the above derivation for the fourth-order temporal derivative approximation is applicable only to the central points within the range $2\tau<=t<=T-2\tau$, where sufficient neighboring points exist to compute the derivative. For temporal boundary points ($t=0, \tau,T-\tau, T$), the formula in Eq.\ref{Eq:temporal_central} cannot be applied directly. As such, we derive specialized formulas for these boundary points by leveraging Taylor expansions with available neighboring data and canceling out higher-order terms in a manner similar to the derivation of Eq. \ref{Eq:temporal_central}, ensuring good accuracy across the entire temporal domain. We utilize the HSP values at $t=0,\tau,2\tau,3\tau,4\tau$ for calculating the first two boundary derivatives $u^\prime_0$ and $u^\prime_\tau$, and the HSP at the last five temporal points $t=T-4\tau,T-3\tau,T-2\tau,T-\tau,T$ to compute $u_{T-\tau}^\prime$ and $u_{T}^\prime$, we can obtain the boundary temporal derivative as: 


\begin{align}
    \frac{\hat{u}_{0}}{\partial t} &\approx \frac{-25\hat{u}_{0} + 48\hat{u}_{\tau} - 36\hat{u}_{2\tau} + 16\hat{u}_{3\tau} - 3\hat{u}_{4\tau}}{12\tau} \nonumber\\
    \frac{\hat{u}_{\tau}}{\partial t} &\approx \frac{-3\hat{u}_{0} - 10\hat{u}_{\tau} + 18\hat{u}_{2\tau} - 6\hat{u}_{3\tau} + \hat{u}_{4\tau}}{12\tau} \nonumber\\
     \frac{\hat{u}_{T-\tau}}{\partial t}&\approx \frac{-\hat{u}_{T-4\tau} + 6\hat{u}_{T-3\tau} - 18\hat{u}_{T-2\tau} + 10\hat{u}_{T-\tau} + 3\hat{u}_{T}}{12\tau}\nonumber\\
     \frac{\hat{u}_{T}}{\partial t} &\approx \frac{3\hat{u}_{T-4\tau} - 16\hat{u}_{T-3\tau} + 36\hat{u}_{T-2\tau} - 48\hat{u}_{T-\tau} + 25\hat{u}_{T}}{12\tau}
\end{align}
Similarly, this setting can be applied to obtain the ND-based temporal derivative for $\hat{v}$, i.e., $\frac{\partial \hat{v}}{\partial t}$.

\end{itemize}

By replacing traditional automatic differentiation with spatiotemporal numerical differentiation, we impose a tighter EP constraint on local spatiotemporal regions rather than simply fitting individual points to satisfy the EP rule. This approach reduces overfitting to the EP loss $\mathcal{L}_{EP}$, ensures the EP laws are effectively applied across the entire domain, and alleviates the computational complexity associated with requiring a large number of collocation points. 
Furthermore, it provides a robust framework for scenarios where the model input lacks explicit spatiotemporal instances, rendering the calculation of derivatives through AD infeasible, while still allowing the EP constraint.

\section{Experimental Design and Results}

The performance of the proposed EAND-ARN is evaluated in a body-heart electrical conduction system. The body and heart geometry is obtained from the 2007 PhysioNet Computing in Cardiology Challenge \cite{goldberger2000physiobank, dawoud2008using}. The heart geometry comprises 1094 nodes and 2184 triangular elements, while the torso surface consists of 352 nodes and 677 triangular elements. The transfer matrix $\boldsymbol{R}$ that characterize the signal transformation between two body and heart surfaces can be derived given the body-heart geometry using Green's second identity \cite{barr1977relating}. The ground truth HSP and BSPM data are obtained based on the simulation of a cardiac reaction-diffusion system \cite{wang2009physiological}. The reference HSP data $u$ consists of $1094 \times 661$ data points, representing 1094 time-series signals defined on the heart geometry, with a temporal resolution of 661 discrete time steps in this study. The BSPM data $y$ has dimensions of $352\times661$, corresponding to 352 time-series signals collected from the body surface. This BSPM data is utilized in the data-driven loss ($\mathcal{L}_D$) to train the EAND-ARN model and ensure the HSP prediction aligns with BSPM observation.

\begin{figure}[!t]
\centering
\includegraphics[width=3.5in]{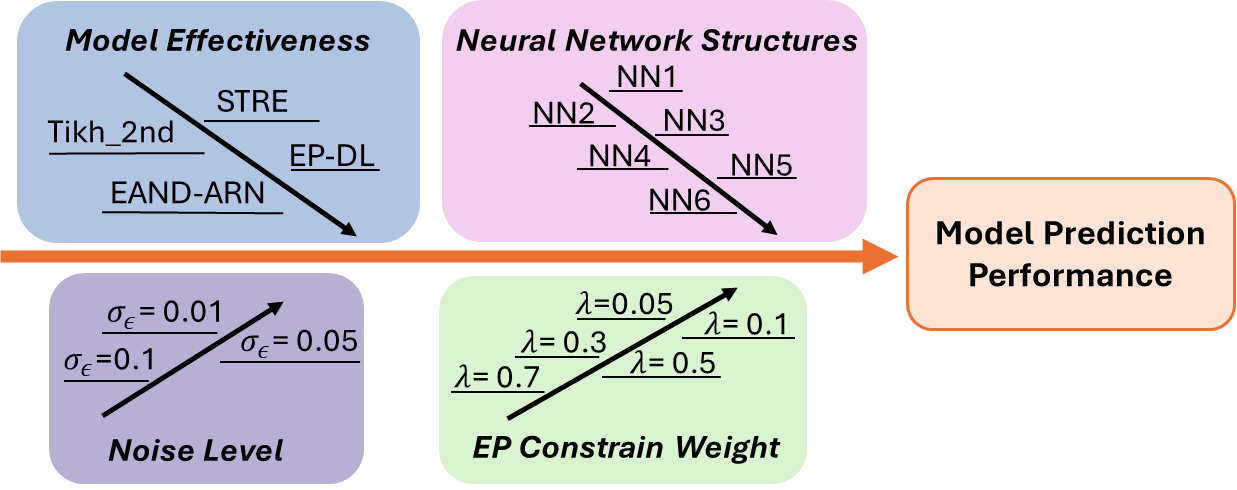}
\caption{Experimental design for model performance evaluation.}
\label{fig:fishbone}
\end{figure}

We assess the effectiveness of the EAND-ARN by comparing its predictive performance in HSP reconstruction against other state-of-the-art methods, including Tikhonov second-order (Tikh\_2nd), spatiotemporal regularization (STRE), and our previous EP-DL model. Additionally, we analyzed the impact of different DNN structures, varying levels of body sensor noise, and EP constraint weights on the HSP reconstruction accuracy. The experimental design is illustrated in Fig. \ref{fig:fishbone}. The model performance will be assessed using three metrics: Relative Error ($RE$), Correlation Coefficient ($CC$), and Mean Squared Error ($MSE$), which are defined as follows:


\begin{align}
	RE &= \frac{\sqrt{\sum_{s,t} \left\|\hat{u}(s,t) - u(s,t)\right\|^2}}{\sqrt{\sum_{s,t} \left\|u(s,t)\right\|^2}}\\
	CC &= \frac{ \sum_s (\hat{u}(s,\cdot) - \bar{\hat{u}}(s,\cdot))^T (u(s,\cdot) - \bar{u}(s,\cdot)) }{ \sqrt{ \sum_s \left\|\hat{u}(s,\cdot) - \bar{\hat{u}}(s,\cdot)\right\|^2 \sum_s \left\|u(s,\cdot) - \bar{u}(s,\cdot)\right\|^2} }\\
	MSE &= \frac{1}{N} \sum_{s,t} \left\|\hat{u}(s,t) - u(s,t)\right\|^2
\end{align}


where $N$ is the total number of the spatiotemporal instances for the heart system. $u(s,t)$ and $\hat{u}(s,t)$ denote the reference and estimated HSP signals, while the overbars indicate the mean of the corresponding time series at node $s$. $RE$ quantifies the deviation of the estimated signal from the reference, with lower values indicating greater accuracy. $CC$ evaluates the similarity between the reference and predicted patterns, where a value of 1 indicates a perfect match and 0 indicates no similarity. $MSE$ reflects the average squared difference between the reference and estimated signals. To evaluate the robustness of the proposed method under noisy conditions, note that a Gaussian noise $ \epsilon(s,t) \sim \mathcal{N}(0, \sigma^2) $ with standard deviation $ \sigma = 0.01 $ is added to the simulated BSPM data $ y(s,t) $, unless otherwise specified.


\subsection{Effectiveness of Spatiotemporal Numerical Differentiation }



We evaluate the effectiveness of the proposed spatiotemporal numerical differentiation (STND) scheme in enforcing the EP rule on the predicted HSP $\hat{u}$. To do this, we employ a feed-forward neural network (FNN) architecture that used in our earlier EP-DL model. However, instead of the 5-layer network in the earlier work \cite{xie2022physics}, we utilize a deeper network with 10 layers, each consisting of 10 neurons, to enhance the presentation of results. Further comparisons of neural network structures are presented in Section \ref{section: impact nerualnetworkstructure}. 
Note that, to ensure reproducibility, each experiment is repeated 10 times, and each metric is reported with its corresponding error bar, representing one standard deviation. To further analyze the impact of spatial and temporal numerical differentiation (ND) in constructing the EP constraint, we compare different configurations of EP-DLs. The EP-DL employing AD for calculating both the spatial Laplacian and temporal derivatives in the EP loss is termed $\text{EP-DL}_{AD}$. The architecture that employs ND for the spatial Laplacian and AD for the temporal derivative in the EP loss is referred to as $\text{EP-DL}_{\Delta_{ND}}$. Meanwhile, the architecture using ND for calculating both the spatial Laplacian and temporal derivatives is referred to as $\text{EP-DL}_{ND}$.

\begin{figure*}[htp]
\centering
\includegraphics[width=7in]{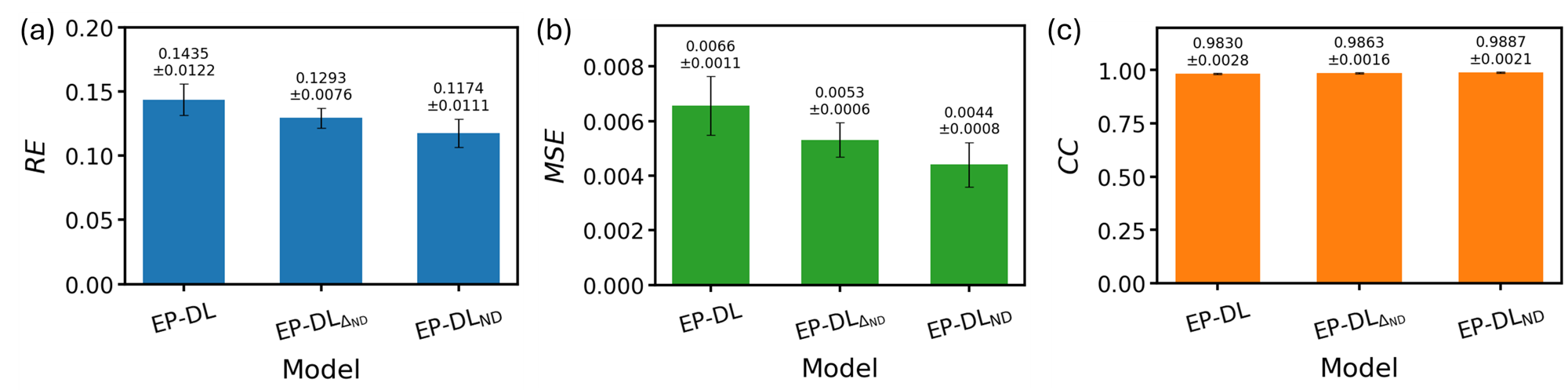}
\caption{(a) $RE$, (b) $MSE$, and (c) $CC$ yielded by EP-DL, $\text{EP-DL}_{\Delta_{ND}}$, and $\text{EP-DL}_{ND}$.}
\label{fig:exp1}
\end{figure*}


The results in Fig. \ref{fig:exp1} (a-c) compare $RE$, $MSE$, and $CC$ across different models. The $\text{EP-DL}_{AD}$ generates the highest $RE$ ($0.1435 \pm 0.0122$), $MSE$ ($0.0066 \pm 0.0011$), and lowest $CC$ ($0.9830 \pm 0.0028$), indicating non-ideal performance. The $\text{EP-DL}_{\Delta_{ND}}$, which incorporates ND for the spatial Laplacian, demonstrates improved accuracy across all three metrics, highlighting the effectiveness of employing ND for 3D irregular geometries, to enhance EP enforcement. The $\text{EP-DL}_{ND}$, which applies ND for both spatial and temporal domains, achieves the best overall performance: $RE$ of $0.1174 \pm 0.0111$, $MSE$ of $0.0044 \pm 0.0008$, and $CC$ of $0.9887 \pm 0.0021$. The difference in performance between $\text{EP-DL}_{\Delta_{ND}}$ and $\text{EP-DL}_{ND}$ suggests that the incorporation of fourth-order Taylor expansion-based ND for the temporal domain further enhances model accuracy and HSP prediction. 
Thus, we adopt the STND scheme for the remainder of our experiments.

\subsection{Impact of Neural Network Structure on the Model Performance}
\label{section: impact nerualnetworkstructure}


 In this subsection, we explore the impact of the neuron numbers per layer on the predictive performance. Building on the selected number of neurons, we further investigate the configuration of the adaptive residual network (ARN). By effectively deepening the neural network while suppressing issues such as ineffective scaling and degradation, we aim to enhance the network's expressiveness and improve the accuracy of HSP predictions.

\textit{1) Selection of Neuron Number:} Expanding the width of the neural network allows it to capture more complex patterns, however, it might complex the optimization, potentially leading to slow convergence. In order to find a proper number of neurons per layer for the subsequential experiment, we fix the number of network layers to be 10, and tested three different options for neuron numbers, with 5, 10, and 15 neurons per layer. 

\begin{figure*}[htp]
\centering
\includegraphics[width=7in]{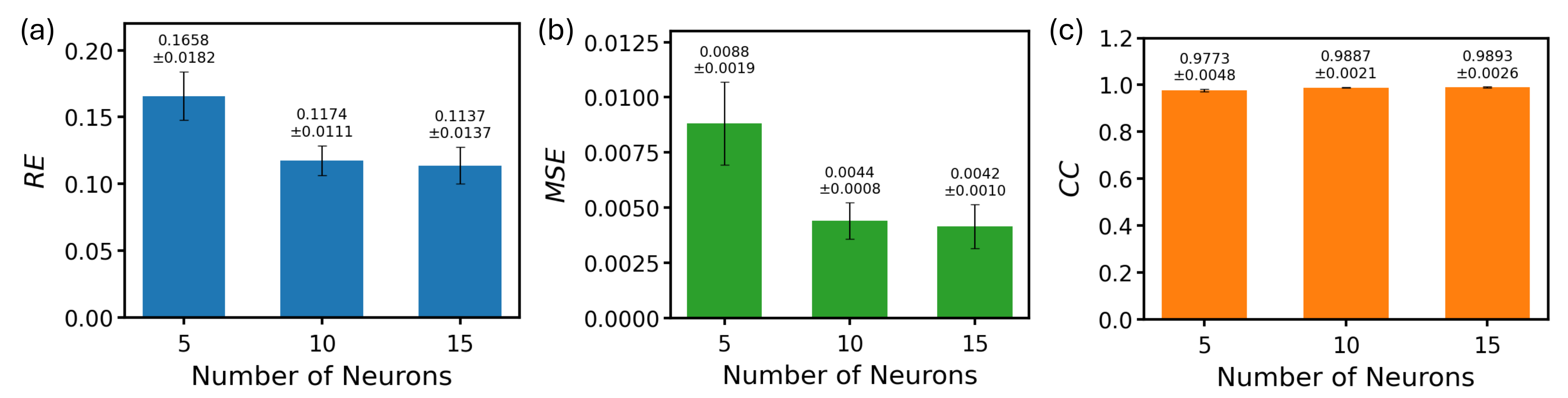}
\caption{The influence of number of neurons on (a) $RE$, (b) $MSE$, and (c) $CC$, for neural networks with 10 layers.}
\label{fig:exp2}
\end{figure*}

As illustrated in Fig. \ref{fig:exp2} (a-c), increasing the number of neurons per layer significantly enhances model performance on all metrics. With five neurons, the model exhibits relatively high error rates, with a $RE$ of $0.1658 \pm 0.0182$ and $MSE$ of $0.0088 \pm 0.0019$. As the neuron number increases to 10 and 15, we observe a continuous improvement in all performance indicators. The 15-neuron configuration achieves the best results, with the lowest $RE$ of $0.1137 \pm 0.0137$ and $MSE$ of $0.0042 \pm 0.0010$, while highest $CC$ $0.9893 \pm 0.0026$. These improvements in $RE$, $MSE$, and $CC$ indicate that a higher number of neurons enables the model to better capture the inverse relationship. However, the marginal gains observed between the ten neurons and fifteen neurons suggest a potential plateau in performance improvement with further complexity increases. 
As such, we choose the neuron number as 15 per layer for subsequent experiments.

\textit{2) ARN Configuration:} 
To evaluate the impact of adaptive residual blocks on predictive performance, we conduct experiments by varying the number of Adap-ResBlocks. The specific details are shown in Table \ref{tab:nnconfig}. These configurations varied in the total number of layers and the number of Adap-ResBlocks. Note that when counting the total number of layers, the layers within each Adap-ResBlock are included in the count. In other words, we are not treating an Adap-ResBlock as a single unit or layer. This is to ensure that comparisons across configurations maintain similar computational complexity. 

 NN1 through NN6 utilize STND for enforcing the EP rule but differ in their neural network structures. NN1 and NN2 consist of 4 layers, while NN3 and NN4 have 7 layers, and NN5 and NN6 have 10 layers, with each layer containing 15 neurons. NN1, NN3 and NN5 use a standard feedforward network structure, whereas NN2, NN4 and NN6 incorporate adaptive residual networks. Specifically, NN2 includes one Adap-ResBlocks, NN4 includes two Adap-ResBlocks, and NN6 includes three Adap-ResBlocks. Each Adap-ResBlock consists of two layers as its functional block, with one additional layer placed between two blocks, as well as at the beginning and end of the network.
This setting enabling us to isolate the effect of residual blocks while controlling for network depth. It allowed us to investigate the effectiveness of residual blocks versus traditional layered structures of equivalent depth. 

\begin{table}[thb]
    \caption{Neural network configurations\label{tab:nnconfig}}
    \centering
    \begin{tabular}{c|c|c|c|c|c|c}
        \hline
        & NN1 & NN2 & NN3 & NN4 & NN5 & NN6 \\ \hline
        Layers & 4 & 4 & 7 & 7 & 10 & 10 \\ \hline
        Neurons & 15 & 15 & 15 & 15 & 15 & 15 \\ \hline
        Residual Blocks & 0 & 1 & 0 & 2 & 0 & 3 \\ \hline
    \end{tabular}
\end{table}

\begin{figure*}[!t]
\centering
\includegraphics[width=7in]{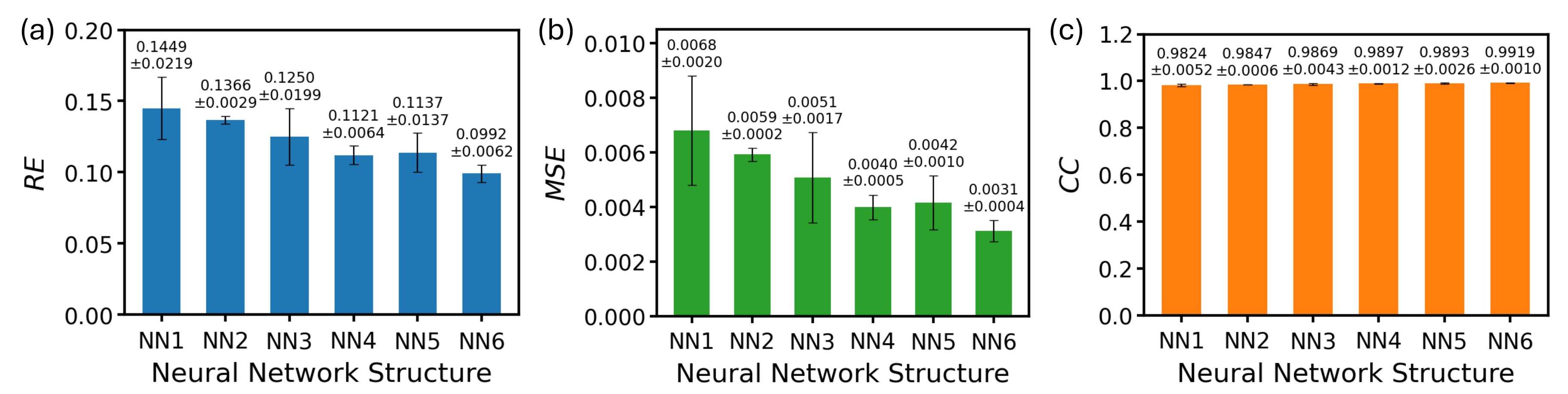}
\caption{The influence of different neural network structures on (a) RE, (b) MSE, and (c) CC.}
\label{fig:exp3}
\end{figure*}

Fig. \ref{fig:exp3} (a-c) illustrates the performance of these configurations on the three metrics. The results demonstrate a clear trend of improvement as we introduce and increase the number of adaptive residual blocks. 
Comparing NN1 and NN2, which have the same total number of layers but differ in the presence of residual blocks, we observe that NN2 with one residual blocks outperforms NN1 in all metrics ($RE$: 0.1366 ± 0.0029 vs $0.1449 \pm 0.0219$ , $MSE$: 0.0059 ± 0.0002 vs 0.0068 ± 0.0020, $CC$: 0.9847 ± 0.0006 vs 0.9824 ± 0.0052). NN4 with two residual blocks also shows superior performance compared to NN3 with the same total layers but no residual blocks ($RE$: $0.1121 \pm 0.0064$ vs 0.1250 ± 0.0199, $MSE$: 0.0040 ± 0.0005 vs 0.0051 ± 0.0017, $CC$: 0.9897 ± 0.0012 vs 0.9869 ± 0.0043). And similarly, NN6 outperforms NN5 ($RE$: 0.0992 ± 0.0062 vs 0.1137 ± 0.0137, $MSE$: 0.0031 ± 0.0004 vs 0.0042 ± 0.0010, $CC$: 0.9919 ± 0.0010 vs 0.9893 ± 0.0026). As increasing the number of residual blocks from one to three, we observe a further consistent decrease in $RE$ and $MSE$, coupled with an increase in $CC$. Notably, NN6 achieves the best performance across all metrics. 
The consistent trend across all three metrics indicates that the introduction of adaptive residual blocks enables the model to learn richer representations, effectively mitigating the ineffective scaling issue commonly observed in traditional FNNs when solving inverse ECG problems. 

Another challenge in training EP-aware deep learning models is that improper weight initialization can cause the EP-aware loss to start at an excessively high value relative to the data-driven loss, leading to a biased training process. The incorporation of Adap-ResBlocks effectively mitigates this bad initialization issue while enabling the network to achieve greater depth. To evaluate the effectiveness of adaptive residual blocks in mitigating the bad initialization problem, we also compare the initialization performance across six neural network configurations. We define bad initialization as an imbalance in the contributions of the data-driven loss and EP-aware loss, where one term dominates the other.  To ensure statistical significance, we ran each experiment 30 times. 
Table \ref{tab:badinitial} presents the results of this experiment. Notably, the configurations with adaptive residual blocks (NN2, NN4, and NN6) demonstrated significantly lower bad initialization rates compared to their counterparts without residual blocks. Specifically, NN2 that contains one Adap-Resblock presents significantly lower bad initial rate than NN1 (10\% vs 60\%). 
NN4, with two Adap-Resblocks, reduced the bad initialization rate to 13.33\% from 76.67\% in NN3, while NN6, with three residual blocks, achieved a 20.00\% rate compared to 83.33\% in NN5. These results strongly indicate that adaptive residual blocks can substantially improve the initialization stability of EP-aware deep learning models, and meanwhile enhance the expandability of the network structure. 

\begin{table}[hb]
    \caption{Comparison of bad initialization rates across different neural network models \label{tab:badinitial}}
    \centering
    \begin{tabular}    {
    			>{\centering\arraybackslash}p{2cm}|
    			>{\centering\arraybackslash}p{0.70cm}|
    			>{\centering\arraybackslash}p{0.70cm}|
    			>{\centering\arraybackslash}p{0.70cm}|
    			>{\centering\arraybackslash}p{0.70cm}|
    			>{\centering\arraybackslash}p{0.70cm}|
    			>{\centering\arraybackslash}p{0.70cm}
    		}
        \hline
        & NN1 & NN2 & NN3 & NN4 & NN5 & NN6 \\ \hline
        Bad Initialization & 18 & 3 & 23 & 4 & 25 & 6 \\ \hline
        Bad Initial Rate & 60.00\% & 10.00\% & 76.67\% & 13.33\% & 83.33\% & 20.00\% \\ \hline
    \end{tabular}
\end{table}

The results above validate the effectiveness of Adap-ResBlock in enhancing predictive capability by increasing the network depth and improving training robustness through the mitigation of poor initialization issues. 
According to the result, we selected NN6 – consisting of 10 layers, 15 neurons per layer, and 3 adaptive residual blocks – as our optimal network structure for subsequent comparisons in this work.

\subsection{Comparison with SOTA Inverse ECG models}

\begin{figure*}[!t]
\centering
\includegraphics[width=5.0in]{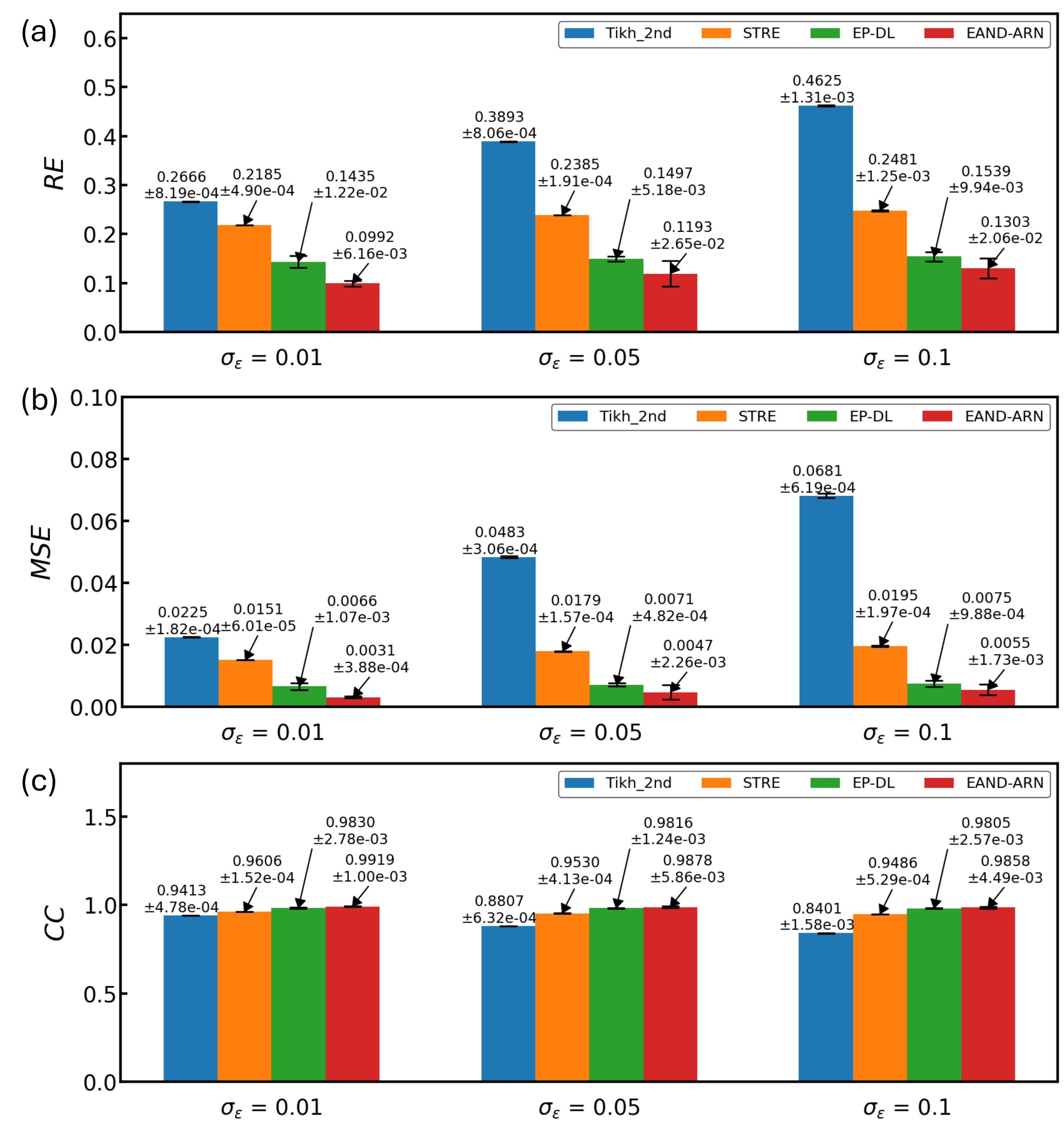}
\caption{The comparison of methods (i.e., Tikh\_2nd, STRE, EP-DL, and EAND-ARN) on (a) RE, (b) MSE, and (c) CC under different BSP noise levels ($\sigma = 0.01, 0.05$, and 0.1)}
\label{fig:exp4}
\end{figure*}

In the current investigation, we conducted a comprehensive comparison study between our proposed EAND-ARN and other SOTA inverse ECG models: Tikh\_2nd, STRE, and EP-DL, which are used as benchmarks. 
In real-world scenarios, noise in BSPM measurements is unavoidable. To examine the impact of measurement noise, we evaluate the proposed model using BSPM data with varying noise levels ($\sigma_{\varepsilon}$ = 0.01, 0.05, and 0.1). Fig. \ref{fig:exp4} presents a detailed comparison of $RE$, $MSE$, and $CC$ for four methods under different noise conditions. 

As the noise level rises from $\sigma =0.01 $ to $\sigma =0.1$, both $RE$ and $MSE$ exhibit a consistent upward trend while $CC$ shows a decreasing trend across all methods. For Tikh\_2nd, $RE$ increases from $0.2666\pm 0.0008$ to $0.4625\pm 0.0013$, $MSE$ rises from $0.0225 \pm 0.0002$ to $0.0681 \pm 0.0006$, and $CC$ decrease from $0.9413\pm 0.0005$ to $0.8401 \pm 0.0016$. Similarly, STRE sees an increase in $RE$ from $0.2185\pm 0.0005$ to $0.2481 \pm 0.0013$, and $CC$ drops from $0.9606 \pm 0.0002$ to $0.9486 \pm 0.0005$. For our previous EP-DL model, $RE$ grows from $0.1435\pm 0.0122$ to $0.1585\pm 0.0117$, $MSE$ increases from $0.0066\pm 0.0010$ to $0.0080 \pm 0.0012$, while $CC$ is reduced from $0.9816\pm 0.0028$ to $0.9793\pm 0.0031$. The proposed EAND-ARN shares the same metrics trend when increasing the noise level--with $RE$ increases from $0.0992 \pm 0.0062$ to $0.1303 \pm 0.0206$, $MSE$ grows from $0.0031$ to $0.0055\pm 0.0017$, and $CC$ declines from $0.9919\pm 0.0010$ to $0.9859 \pm 0.0045$. 
Even as the noise level increases, the proposed EAND-ARN maintains minimal discrepancy with the ground truth HSP dynamics, achieving the lowest 
$RE$ and $MSE$ values and the highest $CC$ value among all models.

\begin{figure*}[htbp]
\centering
\includegraphics[width=7in]{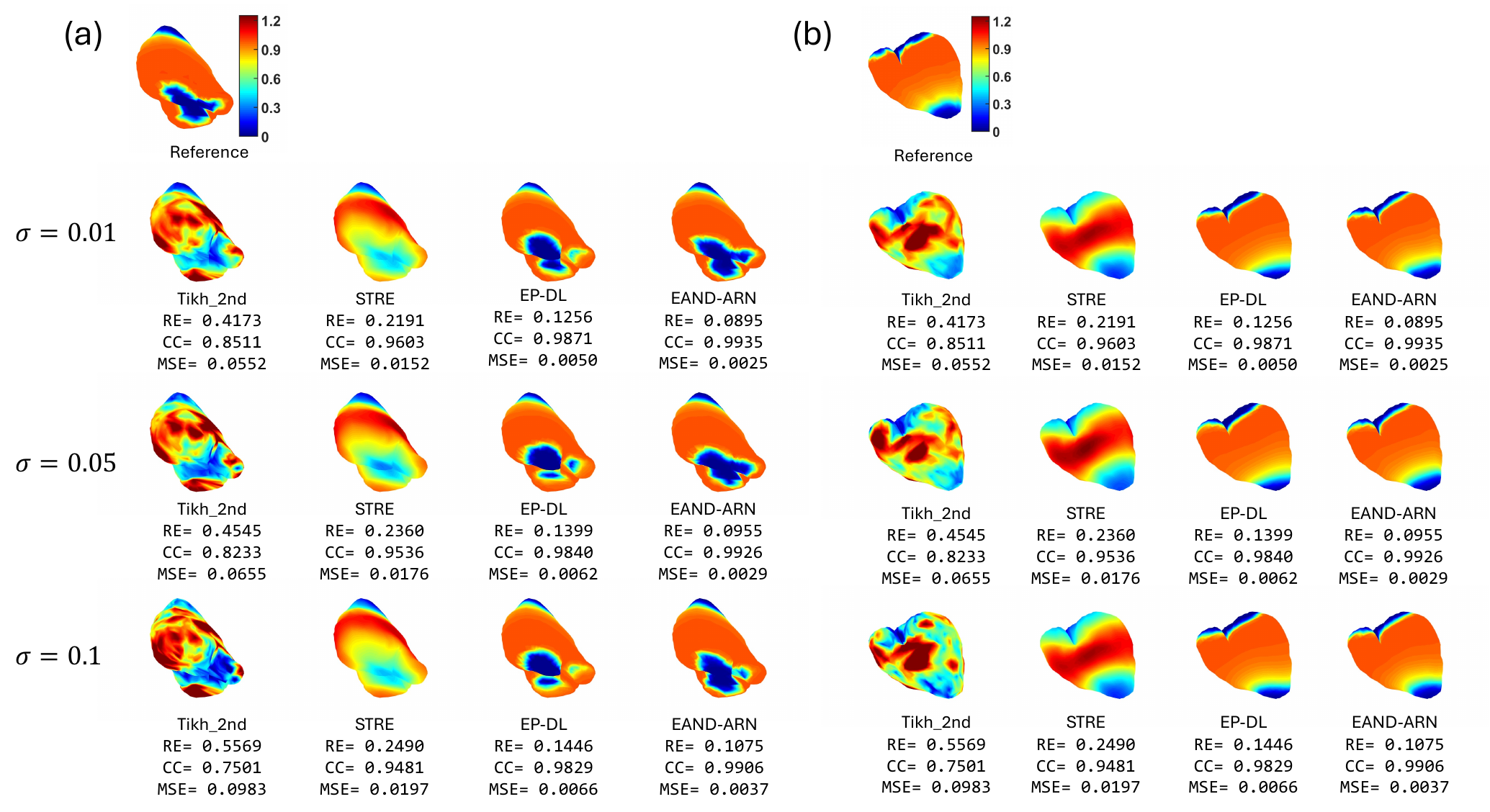}
\caption{Spatial visualization of reference HSP $u$, and the predicted HSP $\hat{u}$ yielded by methods including Tikh\_2nd, STRE, EP-DL, and EAND-ARN with different observation noise level $\sigma = 0.01, 0.05, 0.1$; (a) the basal view; (b) the anterior view.}
\label{fig:exp5}
\end{figure*}

Fig. \ref{fig:exp5}(a-b) show the spatial visualization of the estimated HSP distributions on the heart surface at two different views at one specific time instance, illustrating a depolarization process from the apex to the base plane. In both views, the reference mapping represents the actual potential distribution, while different rows depict the predicted HSP distributions given BSPM observation with varying noise levels ($\sigma= 0.01, 0.05, 0.1$). We select the best run among all executions for the color map presentation. Note that the HSP $u(s,t)$ is normalized, and both 
$u$ and $t$ are dimensionless. The quantitative metrics are noted below the color maps. Tikh\_2nd, STRE, and EP-DL show more pronounced deviations in the performance metrics, especially at higher noise levels. Specifically, at $\sigma_{\varepsilon}$ = 0.1, EAND-ARN achieved the best results with $RE$ of 0.1075, $MSE$ of 0.0037 and $CC$ of 0.9906. In contrast, EP-DL’s performance under the same noise conditions is inferior, with $RE$, $MSE$, and $CC$ values of 0.1446, 0.0066, and 0.9829, respectively. STRE shows a more salient discrepancy with results of 0.2490, 0.0197, and 0.9481, while Tikh\_2nd exhibits the largest errors, with an $RE$ of 0.5569, $MSE$ of 0.0983, and $CC$ of 0.7501. 

Comparing the color patterns of each model to the reference, EAND-ARN consistently provides the closest pattern to the true potential distribution under all noise conditions. 
While the EP-DL model achieves overall good results, the heart map in the basal view (Fig. \ref{fig:exp5}(a)) demonstrates that EAND-ARN provides a more accurate pattern prediction as compared to the reference, particularly in the basal plane region where the geometry is less smooth. In contrast, other regularization-based models, such as Tikh\_2nd and STRE, exhibit significant deviations from the ground truth. Fig. \ref{fig:exp5}(b) illustrates the electrical potential pattern from the anterior view of the heart. The Tikh\_2nd and STRE approaches remain to exibit non-negligible differences compared with the reference in Fig. \ref{fig:exp5}(b). While the EP-DL model produces a smooth signal pattern close to the ground truth, it shows inaccuracies in the apex region, particularly around the iso-potential line on the bottom side, where the HSP exhibits a skewed morphology. In contrast, the EAND-ARN demonstrates the closest match to the reference map. Both the quantitative metrics and spatial visualizations exhibit that our model achieves the highest prediction fidelity and effectively preserves spatial detail in high-noise scenarios, particularly in geometrically complex regions.


\begin{figure*}[!t]
\centering
\includegraphics[width=6in]{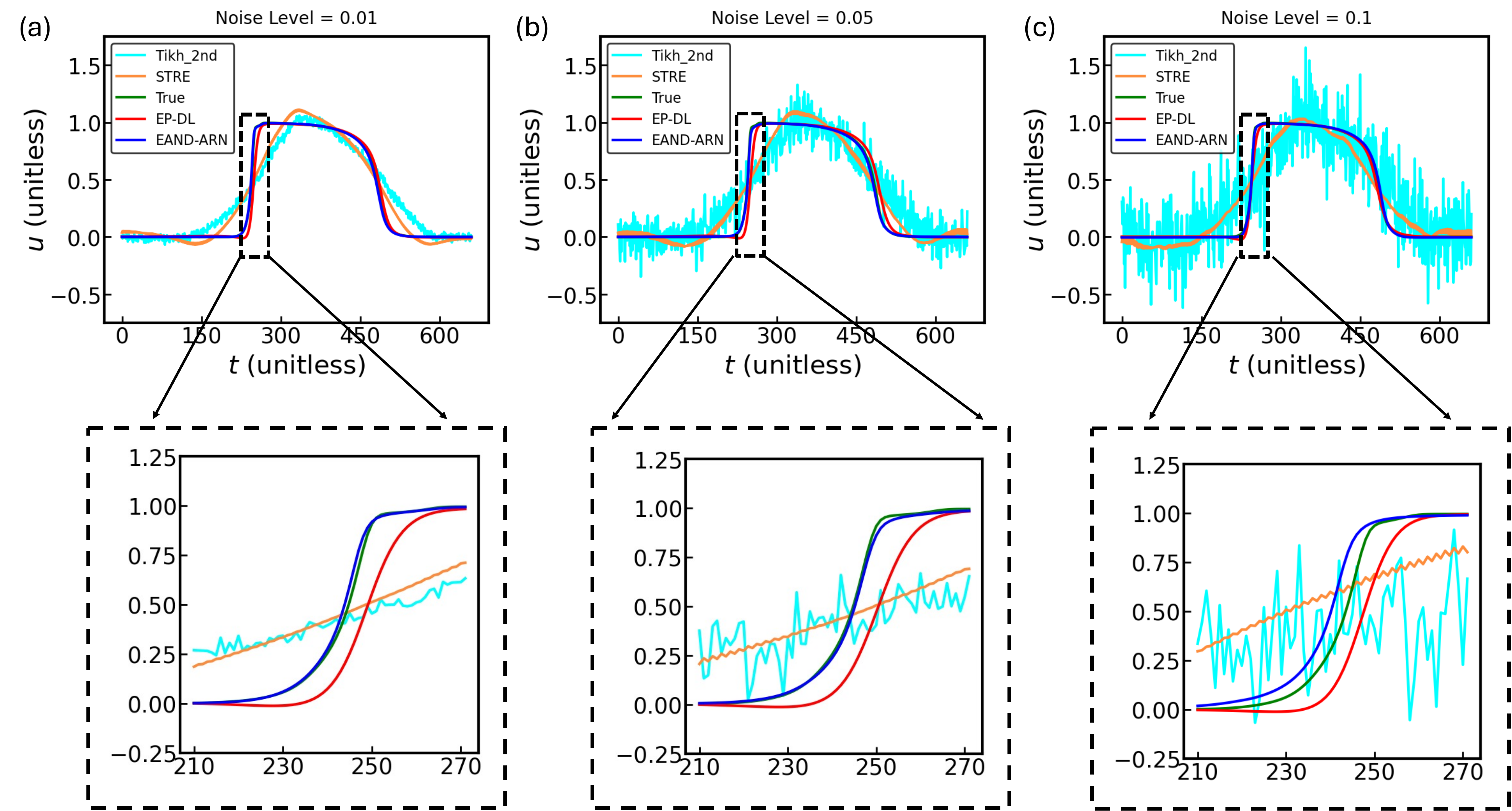}
\caption{ The comparison of estimated HSP evolvement over time at a specific spatial location with noise level (a) 0.01, (b) 0.05, and (c) 0.1}
\label{fig:exp7}
\end{figure*}

Fig. \ref{fig:exp7} (a-c) presents the temporal evolution of the estimated HSP using four methods at a specific spatial location under varying noise conditions ($\sigma = 0.01$ for Fig.\ref{fig:exp7} (a), 0.05 for (b), and 0.1 for (c)). The results show that DNN-based approaches, i.e., EAND-ARN and EP-DL, achieve high accuracy, producing estimates closely aligned with the true HSP, demonstrating the models' robustness against noise. However, in a zoom-in view (dashed-line enclosed region), the temporal evolution curves highlight that EAND-ARN outperforms the EP-DL, particularly during the upstroke phase of the heartbeat. In contrast, Tikh\_2nd shows substantial noise-induced oscillations and low accuracy in reconstructing the HSP. Although the STRE method delivers a smoother result through temporal regularization, it still deviates significantly from the true HSP. By integrating EP prior knowledge of cardiac electrodynamics, EP-DL and EAND-ARN provide more accurate estimations, with EAND-ARN performing best due to its additional consideration of the local spatiotemporal information and innovation on involving Adap-ResBlocks when conducting inverse modeling.

\subsection{Impact of EP Regularization on the Prediction Performance}
\label{section: lambda}
The EP constraint regularization parameter, represented by $\lambda$ in Eq. \ref{Eq: loss_total}, plays a crucial role in balancing the data-driven loss $\mathcal{L}_{\text{D}}$ with the EP awareness loss $\mathcal{L}_{\text{EP}}$. Our study examined the impact of this hyperparameter on prediction performance by empirically testing $\lambda$ values of 0.05, 0.1, 0.3, 0.5, and 0.7. The corresponding prediction metrics are presented in Table \ref{tab:p}, indicating that varying the intensity of the weight of the EP loss $\lambda$ affects model performance. When $\lambda$ is set to 0.1, the model achieves the lowest $RE$ and $MSE$ of $0.0992 \pm 0.0062$ and $0.0031\pm 0.0004$,  and the highest $CC$ of $0.9919 \pm 0.0010$. 
In contrast, further lower or increase  the $\lambda$ values results in less favorable performance, where all the metrics indicates large predictive error. This suggests that the influence of EP constraints should be carefully calibrated. Hence, we choose $\lambda = 0.1$ to intergrate EP regularization without overwhelming the data-driven learning process.

\begin{table}

    \caption{Model performance with different EP regularization parameter $\lambda$	\label{tab:p}}
    \centering
    \begin{tabular}{c|c|c|c}
        \hline
        $\lambda$ & $RE$ & $CC$ & $MSE$ \\ \hline
        $0.05$ & $0.1012 \pm 0.0089$ & $0.9916 \pm 0.0015$ & $0.0033 \pm 0.0006$ \\ \hline
        $0.1$  & $0.0992 \pm 0.0062$ & $0.9919 \pm 0.0010$ & $0.0031 \pm 0.0004$ \\ \hline
        $0.3$  & $0.1256 \pm 0.0021$ & $0.9871 \pm 0.0004$ & $0.0050 \pm 0.0002$ \\ \hline
        $0.5$  & $0.1252 \pm 0.0035$ & $0.9872 \pm 0.0007$ & $0.0050 \pm 0.0003$ \\ \hline
        $0.7$  & $0.1279 \pm 0.0005$ & $0.9866 \pm 0.0001$ & $0.0052 \pm 0.0001$ \\
    \hline
    \end{tabular}
\end{table}

\section{Conclusion}
In this paper, we introduce a novel EAND-ARN model for estimating cardiac electrodynamics from body sensor observations. To mitigate overfitting in the AP model and enhance the spatiotemporal coherence of HSP predictions, we incorporate spatiotemporal numerical differentiation to compute the spatial Laplacian and temporal derivatives, replacing the traditional automotive differentiation approach.  This scheme reduces the need for a large number of collocation points to enforce EP rules while ensuring prediction coherence within local spatiotemporal regime.  Consequently, it delivers more accurate and robust predictions.
Another key contribution of our work is the integration of adaptive residual blocks into the EP-aware deep learning architecture. This enhancement enables the use of deeper neural networks while lowering the risks of ineffective scaling or network degradation. Moreover, it addresses a critical limitation of the traditional EP-DL model by mitigating the impact of poor initialization, which often leads to increased computational time and resource demands. Our comparative analysis demonstrates that the EAND-ARN framework achieves significantly improved results compared to SOTA approaches, including Tikh\_2nd, STRE, and EP-DL methods. The proposed EAND-ARN framework holds great potential for application in other fields involving complex systems governed by physical laws, including biomedical systems, geophysics, and fluid dynamics.

\bibliographystyle{IEEEtran}
\bibliography{references2}

\newpage

\vspace{11pt}


\vspace{11pt}


\vfill

\end{document}